\documentclass{article}[12pt]
\newcommand{\eqref}[1]{(\ref{#1})}

\def\sech{{\rm sech\,}}

\def\sn{{\rm sn\,}}
\def\cn{{\rm cn\,}}

\def\sin{{\rm sin\,}}
\def\cos{{\rm cos\,}}
\usepackage{array}
\usepackage{graphicx,subfigure}
\usepackage{epsfig}
\begin{document}

\begin{center}
{\Large{\bf Construction of localized atomic wave packets }}\\
S. Sree Ranjani$^\dag$\footnote{ssrsp@uohyd.ernet.in}, P. K. Panigrahi$^\ast$\footnote{prasanta@prl.res.in} and A. K. Kapoor$^\dag$\footnote{akksp@uohyd.ernet.in},\\
$^\dag$ School of Physics, University of Hyderabad, Hyderabad 500 046, India. \\
$^\ast$  Indian Institute of Science Education and Research (IISER) Kolkata, Salt Lake, Kolkata 700 106, India and Physical Research Laboratory, Navrangpura, Ahmedabad 380009, India.
\end{center}

\begin{abstract}
    It is shown that highly localized solitons can be created in lower dimensional Bose-Einstein condensates (BEC), trapped in a regular harmonic trap, by temporally varying the trap frequency. A BEC confined in such a trap can be effectively used to construct a pulsed atomic laser emitting coherent atomic wave packets. It is also shown that one has complete  control over the spatio-temporal dynamics of the solitons. The dynamics of these solitons are compared with those constructed in a BEC where the trap frequency is constant. 
\end{abstract}

\noindent
{\bf Pacs}(03.75.-b, 05.45.Yv, 03.75.Pp)


\noindent
\section{Introduction}

The creation of the Bose-Einstein condensate (BEC) \cite{bose} has opened up an area with a potential for huge technological application. On one side, active research, both theoretical and experimental, is being done to understand the properties of this new phase of matter and on the other side practical applications of the condensate are being developed. The fact that a BEC constitutes coherent matter waves analogous to the coherent light waves, makes it the most suitable candidate for the construction of an atom laser. The first atomic laser was constructed in 1997 in MIT \cite{MOM} and other experimental results were reported in \cite{scin}, \cite{bloch}. An atom laser can be used  to develop accurate atomic clocks \cite{kad}, gyroscopes for ultra precision navigation \cite{lecoq} and will also help in the development of coherent atom optics \cite{pas}, \cite{billy}, atom interferometry \cite{ying}, \cite{fat} and atomic transport \cite{stef}.  The realization of bright solitons and soliton trains \cite{ADJ} -\cite{kon} in a BEC has given further impetus to this line of research.

Solitons are ubiquitous in nature and have a unique property that they do not disperse even after traveling long distances. Optical solitons have been successfully used for the purpose of telecommunications \cite{boris}. A bright soliton is a bunch of atoms moving together and its density is more compared to the background. It has been shown in \cite{carr1} that solitonic pulses can be generated in a BEC in an elongated harmonic trap, by simultaneously changing the scattering length and by making the trap expulsive. Generation of soliton pulses for a pulsed atom laser, using the quantum state transfer property was discussed by Liu {\it et. al.,} \cite{Xi}.
The dynamics of an atom laser were studied in \cite{zhang} - \cite{Nich}

     
      It is a well known fact that the Gross-Petaivskii (GP) equation describes the dynamics of a BEC at the mean field level. An analytical method to solve the GP equation for a BEC trapped in an elongated harmonic trap was presented in \cite{atre}. Here, both the scattering length and the trap frequency were time dependent and the phase of the solitonic solution got related to the trap frequency in a non-trivial way.  This relation was exploited to obtain complete control over the spatio-temporal dynamics of the condensate, in addition to obtaining exact expressions for the solutions. This became possible because this relation provided one with the freedom to vary the trap frequency and also to tune the atomic s-wave scattering length. This in turn allowed one to study the effects of variations in trap frequency on the dynamics of the BEC and also to see how the nonlinearity changes with time. In \cite{sree}, the effect of sudden temporal changes in the trap frequency of an expulsive harmonic trap have been investigated and highly localized solitons were observed in certain cases.
      
      In the present paper, the emphasis is on the variations of the trap frequency of a regular harmonic trap and their effect on the soliton dynamics. It is shown that in a BEC trapped in a regular harmonic trap,  suitable temporal variations of the trap frequency lead to the creation of highly compressed solitons. The trap frequency is varied using an oscillatory potential which asymptotically goes to zero.
When compared to the solitons obtained with constant trap frequency, these solitons are highly localized. From the context of a pulsed atomic laser, highly localized solitons are desirable. Hence, the concentration will be on the generation of localized bright solitons and it will be shown that these solitons are stable for certain ranges of $t$. Therefore, one studies the evolution of the bright solitons in these specific regions only.\\
      
\noindent
\section{Solutions of the NLSE}
   
The dynamics of a dilute Bose gas trapped in a cylindrical harmonic trap with a time dependent
confinement in the $z$ direction given by
$V=m^{\prime}\omega^2_{\bot}(x^2+y^2)/2  +
m^{\prime}\omega^2_0(t)z^2/2$, is governed by the GP equation. This equation is reduced to the quasi-one dimensional nonlinear Schr\"odinger equation (NLSE), with a tight confinement in the $z$ direction $(\omega_{\perp}>>\omega_z)$, using a Guassian trial wave function \cite{atre}, \cite{jack}, \cite{sal},
\begin{equation}
\imath\partial_{t}\psi=-\frac{1}{2}\partial_{zz}\psi+\gamma(t)|\psi|^{2}\psi+\frac{1}{2}M(t)z^2\psi, \label{e1}
\end{equation}
where $\gamma(t)= 2a_s(t)/a_B$, $M(t)=\omega_0^2(t)/\omega^2_{\perp}$
and $a_{\perp}=\hbar/m^{\prime}\omega_{\perp}$. Here, $\gamma(t)$ is the
nonlinear coupling which is attractive if $\gamma(t)<0$ and repulsive if $\gamma(t)>0$. $a_s$ is the scattering length and $a_B$ is the Bohr's radius. Note that $M(t)>0$ corresponds to the regular oscillator and $M(t)<0$ corresponds to the expulsive oscillator. The exact solutions of the NLSE are obtained by using the ansatz solution
\begin{equation}
\psi(z,t)=\sqrt{A(t)}F\{A(t)[z-l(t)]\}\exp[\imath\Phi(z,t)],
\label{e2}
\end{equation}
where
\begin{equation}
\Phi(z,t)= a(t)-\frac{1}{2}c(t)z^2 \label{e3z}
\end{equation}
with $a(t)=a_0+\frac{\lambda -1}{2}\int_0^t
A^2(t^{\prime})dt^{\prime}$, $a_0$ being a constant. Here,  $A(t)$ gives the amplitude and
$l(t)$ gives the location of the center of mass of the soliton respectively.
Substitution of $\psi(z,t)$ in the NLSE gives the elliptic
differential equation for $F$
\begin{equation}
F^{\prime\prime}(T)-\lambda F(T) +2\kappa F^3(T)=0  \label{e3}
\end{equation}
with $\kappa=-\frac{\gamma_0}{A_0}$, $T=A(t)(z-l(t))$ and the
differentiation is with respect to $T$. Solutions of this equation are the Jacobi elliptic functions \cite{han}, namely $\sn(T,m), \cn(T,m)$ etc with elliptic modulus $m$.  Importantly, solitonic solutions arise in the limit $m \rightarrow 1$. Specifically, in the attractive nonlinear regime ($\gamma_0<0$),  $F(T)= \cn(T/\tau_0,m)$ in \eqref{e2} and it
describes a bright soliton train, with $\tau_0^2=-A_0 m /\gamma_0$,  $\lambda=(1-2m)/\tau^2_0$. In the limit $m \rightarrow 1$,  $\cn(T,m)   \rightarrow \sech(T)$ and the same equation describes a bright soliton 
\begin{equation}
\psi(z,t)=\sqrt{A(t)}\sech(T/\tau_0)\exp[\imath a(t)-\frac{\imath}{2}c(t)z^2],
\label{e41}
\end{equation}
with $\tau_0^2=-A_0  /\gamma_0$. Similarly, the dark soliton trains are obtained with $F(T) = \sn(T/\tau_0,m)$ in \eqref{e2} in the repulsive
nonlinear regime $(\gamma_0 >0)$ with $\tau_0^2=A_0  /\gamma_0$,  $\lambda=-(1+m)/\tau^2_0$. The dark soliton is obtained in the limit $m \rightarrow 1$ as $\sn(T,m) \rightarrow \tanh(T)$ and is  given by
\begin{equation}
\psi(z,t)=\sqrt{A(t)}\tanh(T/\tau_0)\exp[\imath a(t)-\frac{\imath}{2}c(t)z^2].
\label{e41a}
\end{equation}
Substitution of \eqref{e2} in the NLSE \eqref{e1} gives us the following consistency conditions, with $A_0$ and $\gamma_0$ being constants,
\begin{equation}
A(t)=A_0\exp \left(\int_0^t c(t^\prime)dt^{\prime}\right),\,
\gamma(t)=\gamma_0 \frac{A(t)}{A_0}, \label{e6a}
\end{equation}
and
\begin{equation}
\frac{dl(t)}{dt} +c(t)l(t)=0.   \label{e6b}
\end{equation}
which in turn give us expressions for the control parameters. It is also found that $c(t)$ satisfies the Riccati equation
\begin{equation}
\frac{dc(t)}{dt}-c^2(t)=M(t),  \label{e4}
\end{equation}
which can be mapped to a second order differential equation
using the change of variable
\begin{equation}
c(t)=-\frac{d}{dt}\ln[\phi(t)]. \label{e5}   
\end{equation}
The new equation,
\begin{equation}
-\phi^{\prime\prime}(t)-M(t)\phi(t)=0, \label{e6}
\end{equation}
is in the form of the Schr\"odinger equation, if one writes the ratio of the trap frequencies as $M(t)= M_0+V(t)$. Here, $V(t)$ is like a time dependent potential with eigenvalue $M_0 $. This allows one to modify the temporal profile of the trap frequency using potential models for which the Schr\"odinger equation is exactly solvable.  
 Thus, if one knows the solutions
$\phi(t)$ of \eqref{e6}, one can obtain $c(t)$. This when substituted in \eqref{e6a} and \eqref{e6b}, gives the consistency conditions in terms of $\phi(t)$ as follows
\begin{equation}
A(t)=A0\frac{\phi(0)}{\phi(t)}\,,\, \gamma(t)=
\gamma_0\frac{\phi(0)}{\phi(t)}\,,\, l(t)=l_0\frac{\phi(t)}{\phi(0)}, \label{e29}
\end{equation}
where $l_0$ is a constant. Thus one obtains exact expressions for the control parameters. These allow one to write the complete solution of the NLSE using which one can study and control the spatio-temporal dynamics of the solitons. Using \eqref{e4} and \eqref{e5} one can vary the trap frequency and obtain corresponding soliton solutions and study their spatio-temporal dynamics.\\

\noindent
\section{Soliton dynamics}

 As mentioned in the introduction, the present paper looks at the soliton dynamics when the trap frequency of the regular harmonic trap is modelled by an oscillatory function which vanishes as $t \rightarrow 0$. Such a variation in the trap frequency is obtained by taking the deformed free particle potentials \cite{pap} as $V(t)$ in \eqref{e6}. The eigenfunctions of these potentials are known explicitly and are used to obtain the solitonic solutions.
 
\noindent
{\bf Case I} The first potential to be used is  $V(t)\equiv V(t, \lambda)$, $t>0$,
\begin{equation}
V(t,\lambda)=\frac{32 M_0\cos^4(\sqrt{M_0}\, t)}{D_0^2} + \frac{8 M_0 \,\sin(2 \sqrt{M_0}\, t)}{D_0}, \label{e12}
\end{equation}
where $D_0 = 2\sqrt{M_0}\,t +\sin (2\sqrt{M_0}\,t)+4\sqrt{M_0}\,\lambda$, $\sqrt{M_0}>0$ and $ \lambda > 0$. The square integrable
solution for \eqref{e6} with the above $V(t)$ is
\begin{equation}
\tilde{\phi}_1(t, \lambda) = \frac{4\sqrt{M_0}\,\cos(\sqrt{M_0}\,t)}{D_0} \label{e13}
\end{equation}
with energy eigenvalue $M_0$. Using the above solution in \eqref{e29}, one
obtains
\begin{equation}
A(t) = A_0 \left[\frac{D_0 }{ 4 \lambda \sqrt{M_0}}\right]\sec(\sqrt{M_0}\,t), \label{e14}
\end{equation}
\begin{equation}
\gamma(t)=\gamma_0 \left[\frac{D_0 }{ 4\lambda \sqrt{M_0}}\right]\sec(\sqrt{M_0}\,t) \label{e15}
\end{equation}
and
\begin{equation}
l(t)=\left[\frac{4
\lambda \sqrt{M_0}}{D_0}\right]\cos(\sqrt{M_0}\,t).
\label{e16}
\end{equation}
Substitution of the above equations in \eqref{e41} gives the  bright soliton solution. 

The  variation of the nonlinearity with time and the dynamics of the solution as it evolves in space and time are shown in the fig. 1. It can be seen from fig. 1(a) that the nonlinearity changes sign periodically and the change is discontinuous. Since the emphasis is on the study of bright solitons, one chooses a range where the nonlinearity is negative and continuous as shown in fig. 1(b). In this range of $t$, the matter wave density $|\psi(z,t)|^2$, is evolved for varying $z$ and $t$ as shown in fig. 1(c). One can see that the bright soliton is localized and with time gets compressed. The amplitude of the soliton can be controlled by varying $A_0$ in \eqref{e14} and the location of the soliton can be varied by changing $l_0$ in \eqref{e16}.  Thus, one gets to control the dynamics of the soliton completely. For comparison we plot the dynamics of the bright soliton when the trap frequency is constant {\it i.e.} $V(t)=0$. Thus \eqref{e6} is nothing but the Schr\"odinger equation for a free particle. We use the solution $\phi_1(t)=\cos(\sqrt{M_0}\,t)$ and obtain the bright soliton solution and plot its evolution which is shown in fig. 1(d). 

When one compares the dynamics of these two solitons taking the same parameter values, 
one observes that in the earlier case  the soliton is highly localized and compressed whereas in the later case, the soliton moves within the specific $t$ range with more or less the same amplitude, undergoing little or no change. Also one observes huge amplification of the soliton in the case where the trap frequency is time dependent. It has been see that for both the cases, smaller the value of $\sqrt{M_0}$, bigger will be the range of $t$ in which the nonlinearity is continuous, being either attractive or repulsive and hence longer the life time of the soliton. From the above example it is clear that by varying the trap frequency temporally, one can generate solitons which are highly localized and  amplified. Hence, this kind of traps can be used in the construction of pulsed atom lasers.

 In the above case, the solitons have been evolved for $t>0$. One can do a similar analysis for $t<0$ with $\lambda<0$ and it turns out there is no change in the dynamics of the soliton. The only difference lies in the behavior of the nonlinearity plotted in fig. 2(a). One can see that within a certain range of time and with the same parameter values, the nonlinearity changes sign smoothly as shown in fig. 2(b). This implies that within this range the nonlinearity changes its character from being attractive to being repulsive in a smooth way without one resorting to the use of the Feshbach resonance \cite{tie} - \cite{cornish}.  This when compared to the case where the trap frequency is constant, one observes that the regions of attractive and repulsive nonlinearity get interchanged in the two cases and $\gamma(t)$ changes its nature periodically and discontinuously along the negative $t$ axis for the later case. 

\noindent
{\bf Case II }Another example for which the soliton dynamics is studied is when the trap frequency is modulated using 
\begin{equation}
V_2(t,\lambda)=\frac{32 M_0 \sin^4(\sqrt{M_0}\, t)}{D_1^2} - \frac{8 M_0 \sin^2(2 \sqrt{M_0}\, t)}{D_1} \label{e18}
\end{equation}
for $t>0$. Here,  $D_1 = 2\sqrt{M_0}\,t -\sin (2\sqrt{M_0}\,t)+4\sqrt{M_0}\lambda$, $\sqrt{M_0}>0$ and $ \lambda > 0$. This potential is also an oscillating potential which vanishes at infinity. The square integrable
solution of \eqref{e6} with the above $V(t)$ is
\begin{equation}
\tilde{\phi_2}(t, \lambda) = \frac{4\sqrt{M_0}\,\sin(\sqrt{M_0}\,t)}{D_1} \label{e19}
\end{equation}
with eigenvalue $M_0 >0$. Thus using the above solution in \eqref{e29}, one
obtains
\begin{equation}
A(t) = A_0 \frac{D_1}{D^{\prime}_1}\frac{\sin(\sqrt{M_0}\,t_0)}{\sin(\sqrt{M_0}\,t)}, \label{e20}
\end{equation}
\begin{equation}
\gamma(t)=\gamma_0 \frac{D_1}{D^{\prime}_1}\frac{\sin(\sqrt{M_0}\,t_0)}{\sin(\sqrt{M_0}\,t)}\label{e21}
\end{equation}
and
\begin{equation}
l(t)=l_0\frac{D^{\prime}_1}{D_1}\frac{\sin(\sqrt{M_0}\,t)}{\sin(\sqrt{M_0}\,t_0)},
\label{e22}
\end{equation}
where $D^{\prime}_1 = 2\sqrt{M_0}\,t_0 -\sin (2\sqrt{M_0}\,t_0)+4\sqrt{M_0}\lambda$. As in the previous case, the nonlinearity changes sign periodically  and discontinuously.

 Hence, for a certain range of $t$ where $\gamma(t)$ is continuous and negative, the soliton is evolved and the dynamics are shown in fig. 3(a). For comparison the dynamics of a soliton, for the same parameter values, in a BEC located inside a trap with constant trap frequency {\it i.e.,} $(V(t)=0)$ are given in figure 3(b). Here $\phi_2(t)=\sin(\sqrt{M_0}\,t)$ was used to obtain the analytical expression for the bright soliton.  In contrast to case I, both the solitons evolve in  the same way  and have almost the same amplitude except that in the case where the frequency is time dependent, the soliton is more localized. 

\noindent
\section{Conclusions}

In the present paper, it has been shown that in a BEC confined is a regular harmonic trap, a suitable temporal variation of the trap frequency, allows one to construct highly localized and compressed solitons in the scenario where there is no loss or gain of atoms in the condensate. The difference in the soliton dynamics when the trap frequency is varied and when it is constant can be seen very clearly in case I. Thus, by using  suitable oscillatory, asymptotically vanishing potentials to vary the trap frequency, one can construct coherent atomic wave packets in a BEC. These can be used effectively in an atomic laser. Moreover this type of variation in the trap frequency can be implemented with ease in an experimental set up. 
The fact that the solutions of a linear Schr\"odinger equation can be combined with the solutions of the NLSE effectively to obtain analytical solutions of the GP equation allowed one to change the trap frequency. This method also gives one a complete control over the soliton dynamics. Finally one would like to point out here that one can also construct and study the evolution of bright and dark solitonic trains and the dark solitons using the equations \eqref{e41} and \eqref{e41a} in appropriate limits.\\

\noindent
{\bf Acknowledgments}\\
S. S. R. acknowledges financial support provided by Council of Scientific and Industrial
Research (CSIR), Government of India.
\\


\noindent
{\bf References}\\



\begin{figure*}
\centering \subfigure[The nonlinearity $\gamma(t)$]
{\epsfig{file=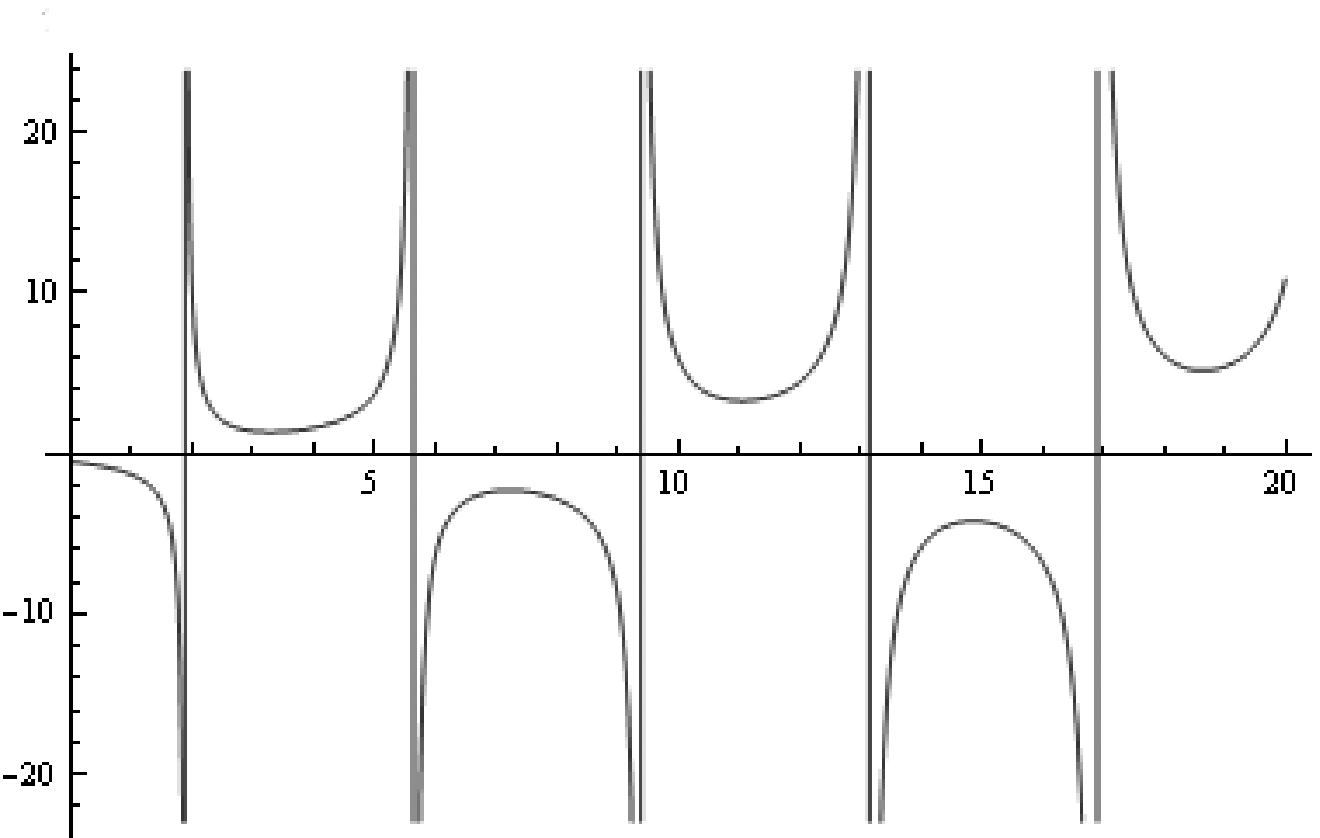,scale=0.4}} \hspace{0.35in}
\subfigure[$\gamma(t)$ in a specific range of $t$: $5.85<t<9.36$]
{\epsfig{file=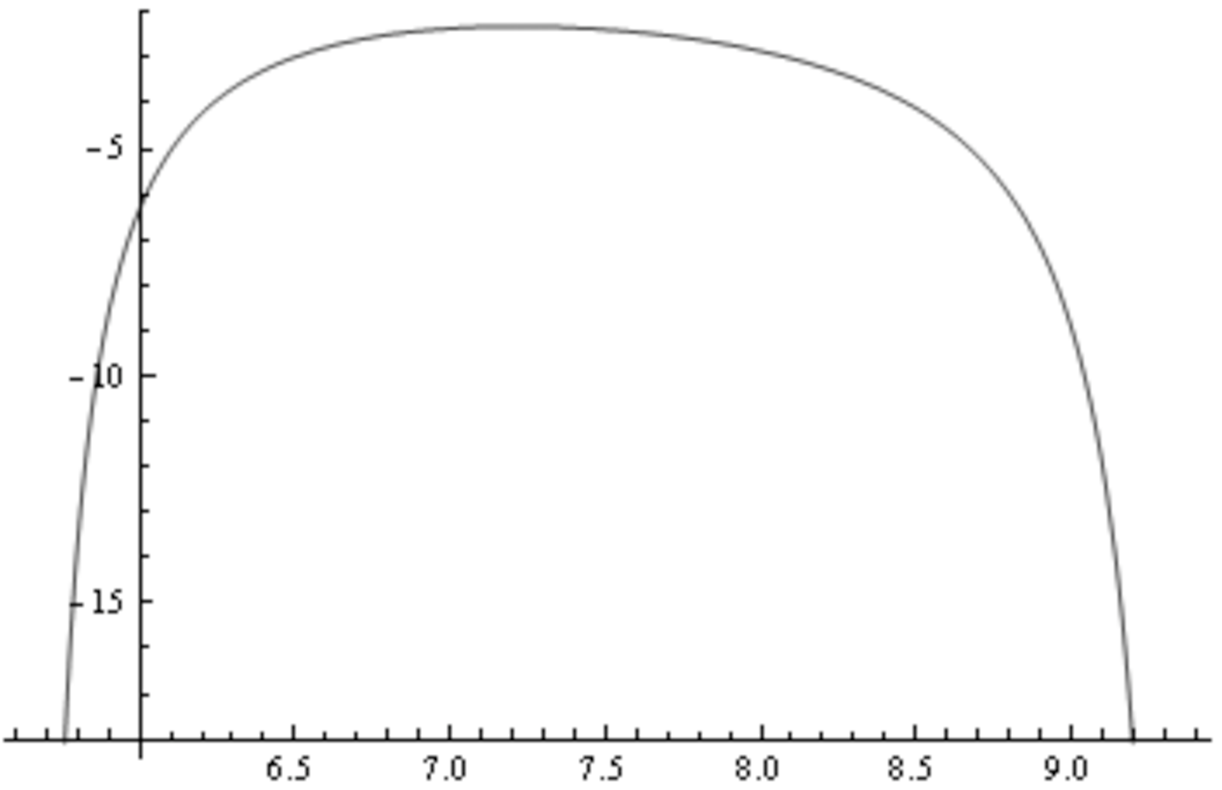,scale=0.4}}
\vspace{0.3in}
\subfigure[Bright soliton in the trap with varying trap frequency]
{\epsfig{file=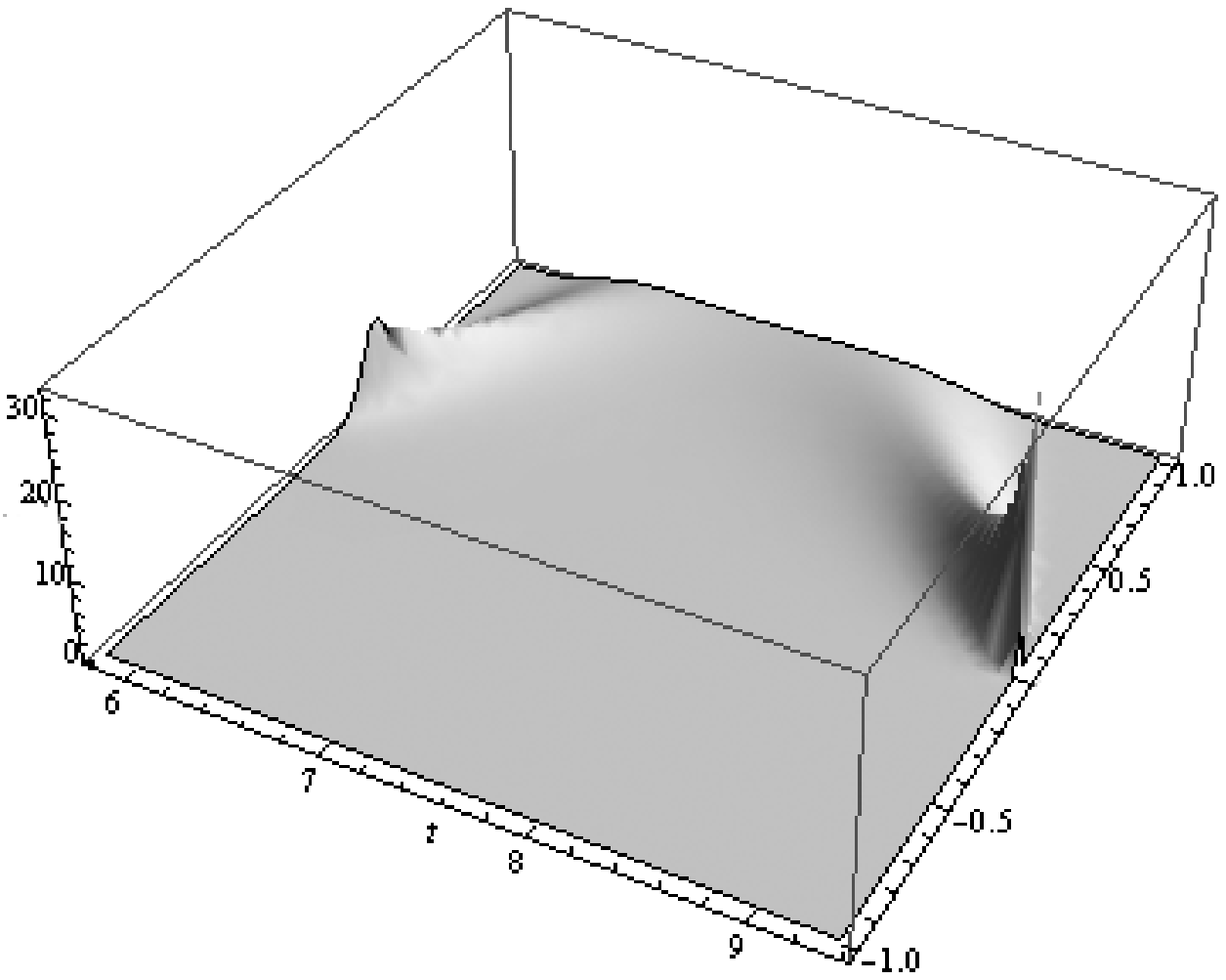,scale=0.4}}\hspace{0.35in}
\subfigure[Bright soliton in the trap with constant trap frequency]
{\epsfig{file=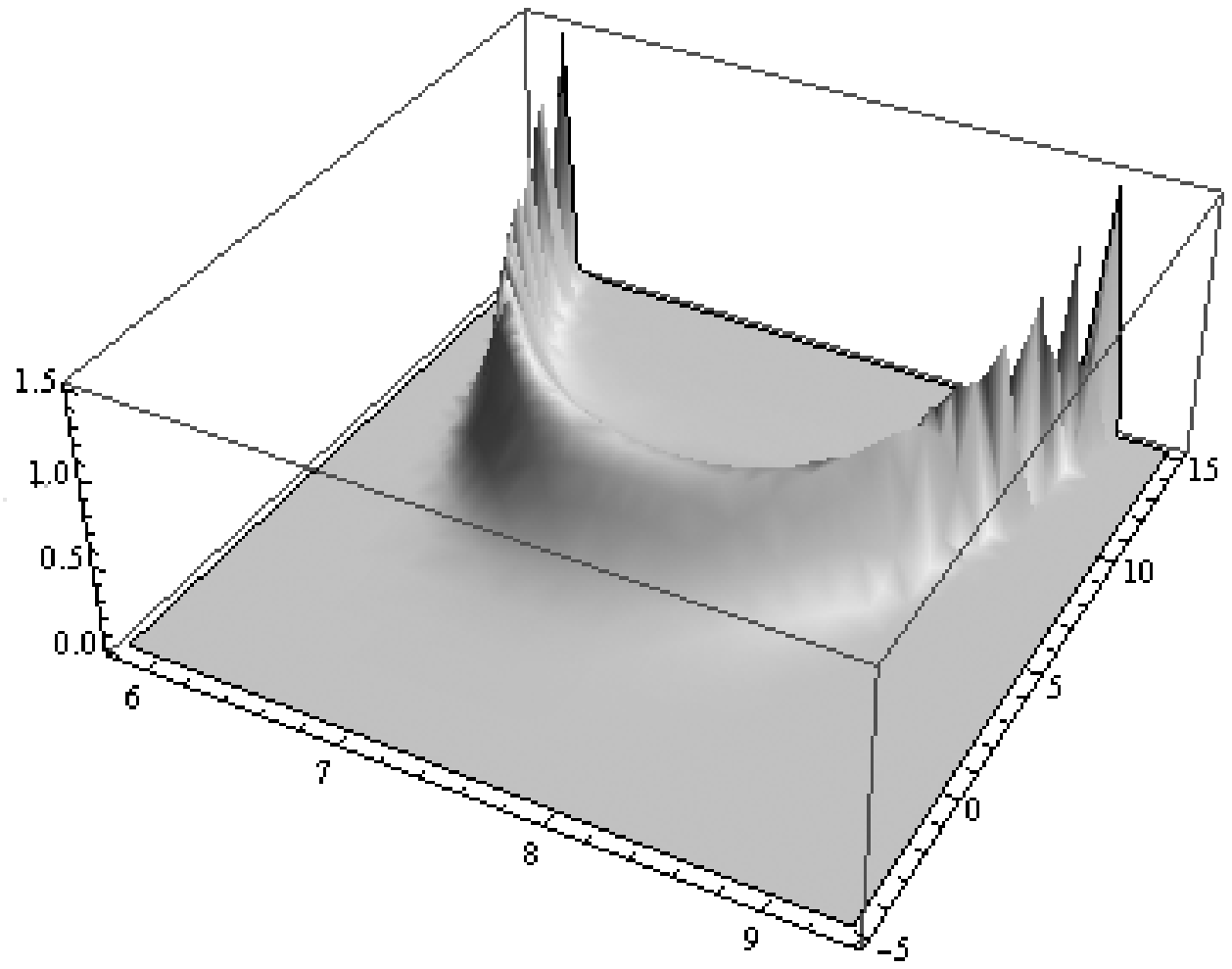,scale=0.4}}
\caption{ The plots 1(a) and 1(b) describe the behaviour of nonlinearity when the trap frquency is time dependent ($\gamma(t)$ (y-axis), $t$ (x-axis)). The plots 1(c) and 1(d) show the evolution of the bright solitons, in the approximate range $t= 5.85$ to $t=9.36$, described by the equation \eqref{e41} ($\left|\psi^2(z,t)\right|$ (z-axis), $t$ (x-axis) and $z$ (y-axis)). The parameter values are $\gamma_0=-0.5$, $M_0=0.83666$, $A_0=1$, $l_0=5$, $\tau_0=1$ and $\lambda=1$. }
\end{figure*}


\begin{figure*}
\centering \subfigure[The nonlinearity $\gamma(t)$]
{\epsfig{file=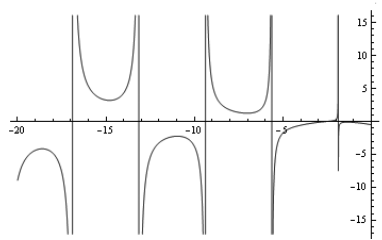,scale=0.4}} \hspace{0.35in} 
\subfigure[The nonlinearity $\gamma(t)$ in a specific range $-5.2<t<-1.5$]
{\epsfig{file=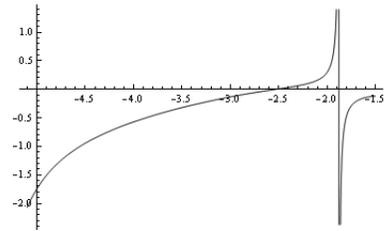,scale=0.4}}
\caption{ The plots of the nonlinearity $\gamma(t)$ in the negative $t$ domain ($\gamma(t)$ (y-axis), $t$ (x-axis)). The parameter values are $\gamma_0=-0.5$, $M_0=0.83666$ and $\lambda=1$.  }
\end{figure*}

\begin{figure*}
\centering \subfigure[Bright soliton in the trap with varying trap frequency]
{\epsfig{file=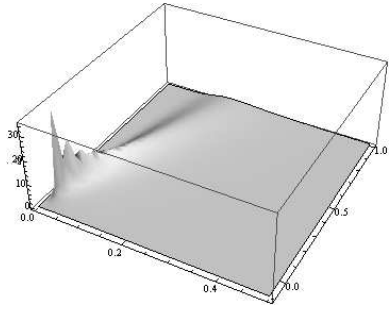,scale=0.4}} \hspace{0.35in} \subfigure[Bright
soliton in the trap with constant trap frequency ]
{\epsfig{file=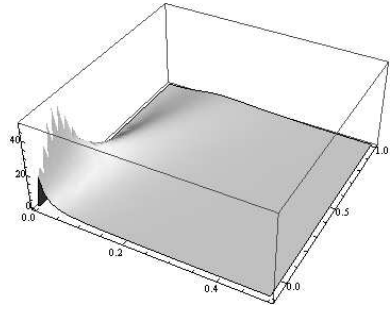,scale=0.4}}
\caption{ The evolution of the bright solitons discussed in case II  ($\left|\psi^2(z,t)\right|$ (z-axis), $t$ (x-axis) and $z$ (y-axis)). The range  of $t$ is $ 0< t < 3.745 $. The parameter values are $\gamma_0=-0.5$, $M_0=0.83666$, $A_0=1$, $l_0=5$, $\tau_0=1$ and $\lambda=1$. }
\end{figure*}

\end{document}